\documentclass[useAMS,usenatbib]{mn2e}
\usepackage{graphicx}
\usepackage{url}

\newcommand{\AlII}{Al\,\textsc{ii}}

\newcommand{\CII}{C\,\textsc{ii}}

\newcommand{\FeII}{Fe\,\textsc{ii}}

\newcommand{\HI}{\textrm{H}\,\textsc{i}}

\newcommand{\HII}{\textrm{H}\,\textsc{ii}}

\newcommand{\Lya}{Ly$\alpha$}

\newcommand{\NHI}{$N(\textrm{H}\,\textsc{i})$}

\newcommand{\NI}{N\,\textsc{i}}
\newcommand{\NII}{N\,\textsc{ii}}

\newcommand{\OI}{O\,\textsc{i}}
\newcommand{\SII}{S\,\textsc{ii}}
\newcommand{\SiII}{Si\,\textsc{ii}}

\def\ltsima{$\; \buildrel < \over \sim \;$}
\def\simlt{\lower.5ex\hbox{\ltsima}}
\def\gtsima{$\; \buildrel > \over \sim \;$}
\def\simgt{\lower.5ex\hbox{\gtsima}}

\title[A carbon-enhanced metal-poor DLA]
{A carbon-enhanced metal-poor damped Ly$\alpha$ system:
Probing gas from Population III nucleosynthesis?\thanks{Based on 
data obtained at the W.~M. Keck Observatory, 
which is operated as a scientific partnership among the 
California Institute of Technology, the University of California, and 
NASA, and was made possible by the generous financial support of the W.~M. Keck Foundation.}}

\author[Cooke et al.]{Ryan Cooke$^{1}$\thanks{email: rcooke@ast.cam.ac.uk}, 
Max Pettini$^{1}$, Charles C. Steidel$^{2}$, Gwen C. Rudie$^{2}$,
\newauthor and Regina A. Jorgenson$^{1}$\\
$^1$Institute of Astronomy, Madingley Road, Cambridge, CB3 0HA\\ 
$^2$California Institute of Technology, MS 249-17, Pasadena, CA 91125, USA}

\begin{document}

\date{Accepted . Received ; in original form }
\pagerange{\pageref{firstpage}--\pageref{lastpage}} 
\pubyear{2010}

\maketitle

\label{firstpage}

\begin{abstract}

We present high resolution observations 
of an extremely metal-poor damped Ly$\alpha$ system,
at $z_{\rm abs} = 2.3400972$ in the spectrum of the QSO
J0035$-$0918,
exhibiting an abundance pattern consistent 
with model predictions for the supernova yields of Population III stars. 
Specifically, this DLA has [Fe/H]\,$\simeq -3$, shows
a clear `odd-even' effect, and is C-rich with
[C/Fe] $= +1.53$, 
a factor of $\sim 20$ greater than reported 
in any other damped Ly$\alpha$ system. 
In analogy to the carbon-enhanced metal-poor 
stars in the Galactic halo (with [C/Fe] $> +1.0$), 
this is the first reported case of a carbon-enhanced damped Ly$\alpha$ system. 
We determine an upper limit to the mass of 
$^{12}$C,  $M(^{12}$C)\,$\leq 200\,\,M_{\odot}$, which depends on 
the unknown gas density $n$(H); if  $n$(H)\,$ > 1$\,cm$^{-3}$
(which is quite likely for this DLA given its low velocity dispersion),
then  $M(^{12}$C)\,$\leq 2\,\,M_{\odot}$, consistent with 
pollution by only a few prior supernovae.
We speculate that DLAs such as the one reported here 
may represent the `missing link'  between
the yields of Pop~III stars and their later incorporation in the
class of carbon-enhanced metal-poor stars which show no enhancement of
neutron-capture elements (CEMP-no stars).

\end{abstract}


%%%%%%%%%%%%%%%%%%%%%%
\section{Introduction}
%%%%%%%%%%%%%%%%%%%%%%

Damped \Lya\ systems (DLAs) are the neutral gas reservoirs 
at high redshift that have, 
by definition, neutral hydrogen column densities in excess of 
$10^{20.3}\, {\rm atoms\,\,cm}^{-2}$ 
(see \citealt{WolGawPro05} for a review). 
At these high column densities the gas is self-shielded \citep[e.g.][]{Vla01}, 
resulting in a simple ionization structure which facilitates 
the derivation of element abundances.
Moreover, the abundances thus derived are independent 
of the geometrical configuration and thermodynamical state
of the gas, and of most other factors  which complicate the analysis
of stellar spectra \citep[e.g.][]{Asp05}. 
The largest uncertainties in DLA abundance studies 
are due to the effects of line saturation and dust depletion (for some elements), 
although the latter of these concerns is found to be minimal 
when the metallicity of the DLA is below $\sim 1/100 \, Z_{\odot}$
\citep[][]{Pet97,ProWol02, Ake05}.

In recent years, these very metal-poor (VMP) DLAs,
with metallicities [Fe/H]\footnote{We adopt the standard 
notation:
${\rm [A/B]} \equiv \log(N_{\rm A}/N_{\rm B}) - \log(N_{\rm A}/N_{\rm B})_{\odot}$
where $N_{\rm A,B}$ refers to the number of atoms of element A,B.}\,$ < -2$, 
have attracted an increasing amount of interest because
of their potential for probing gas which may still bear the chemical imprint
of the first few generations of stars to have formed in the Universe
\citep[e.g.][]{Ern06}.
They are thus an extremely valuable complement, at high redshifts,
to local studies of the oldest and most metal-poor stars in the Galactic halo.

The VMP DLA regime was largely unexplored until very recently, when it
became possible to identify candidate metal-poor DLAs in
Sloan Digital Sky Survey (SDSS) quasars, and then 
measure their chemical composition  with high resolution
follow-up spectroscopy. The first
high spectral resolution (R$\simgt 30000$, FWHM$\simlt 10\,\,{\rm km~s}^{-1}$)
sample of VMP DLAs was compiled by \citet{Pet08},
whose main goal was to compare the relative abundances of C, N, and O
with the trends observed in VMP halo stars in our Galaxy. More
recently, \citet{Pen10} have presented medium resolution 
(FWHM $\sim 60\,{\rm km\,\,s}^{-1}$) spectroscopy of a
sample of 27 VMP DLAs 
to explore the general properties of DLAs in the VMP regime.
However, as Penprase et al.  acknowledge themselves, there are difficulties
with accurately measuring column densities from medium 
(as opposed to high) resolution
data, given that most DLAs
with metallicities less than 1/100 solar exhibit very low velocity
dispersions, with metal line widths less than 10\,km~s$^{-1}$
\citep{Led06,Mur07,Pro08}. 
Under these circumstances, line saturation can easily be overlooked.

Perhaps the most startling result 
from abundance studies of VMP DLAs are the near-solar values of [C/O] 
at low metallicity \citep{Pet08}, in line with the peculiar upturn in 
the [C/O] abundance below [O/H]\,$\simlt -1.0$ in Galactic halo stars
discovered by \citet{Ake04}  and later confirmed
by \citet{Fab09a}. \citet{Ake04} attributed this behaviour to
an increased C yield from earlier generations of massive stars. 
\citet{Pen10} extended this work and reported a number of DLAs with 
supersolar [C/O] suggesting that this ratio continues 
to increase with decreasing [O/H]. 
A more recent compilation of high spectral resolution 
observations of VMP DLAs, however, 
suggests that  the ratio 
in fact plateaus at ${\rm [C/O]}\sim -0.2$ \citep{Coo10b}.

In contrast to the relatively new interest in VMP DLAs, 
studies of metal-poor Galactic halo stars have received 
ongoing attention for more than two decades, 
most recently from the dedicated HK \citep{BeePreShe92} 
and HES \citep{Chr01} surveys. 
A relevant result emerging from this work is 
that nearly one-quarter of all metal-poor stars with [Fe/H]\,$< -2.0$
exhibit a marked carbon enhancement, with [C/Fe]\,$> +1.0$ 
\citep{BeeChr05,Luc06}. 
These are collectively known as carbon-enhanced metal-poor stars (CEMP stars), 
and have been subdivided into four classes based on the abundances 
of their neutron-capture elements: (i \& ii) The CEMP-s and CEMP-r classes, 
with enhancements of elements produced predominantly 
by the $s$-process and $r$-process respectively;
(iii) the CEMP-rs class, with enhancements in both the $s$- and $r$-process elements; and
(iv) the CEMP-no class, which exhibits no such enhancements.
%, with an abundance pattern consistent with that of
%a more `typical' halo star, just having more carbon.

For further details of these classes, and the likely origins of their 
carbon enhancements, we direct the reader to \citet{Mas10}. In short, 
there is reasonable evidence to suggest that CEMP-s and CEMP-rs stars 
are extrinsically polluted by a now extinct asymptotic giant branch 
(AGB) companion. The origin of the CEMP-no class, however, is not yet 
firmly established. Whilst several models invoking mass transfer from 
an AGB companion could explain the lack of neutron-capture elements 
(e.g. \citealt{FujIkeIbe00,SieGorLan04}), radial-velocity measurements 
have not yet confirmed whether CEMP-no stars are host to binary 
companions. Indeed, the apparent difference in the metallicity 
distributions between CEMP-no stars and the other CEMP classes, 
in the sense that CEMP-no stars are more numerous at lower metallicity, 
might suggest that a mechanism other than the AGB binary transfer scenario 
produces the C enhancement in CEMP-no stars \citep{Aok07}.
%However, given that some time must pass before the bulk of the $s$-process 
%elements are released from low mass stars (corresponding to 
%[Fe/H]$\simgt-2.0$, \citealt{Sim04,SneCowGal08}), the distinction 
%between CEMP-s and CEMP-no stars is unclear.
In all likelihood, as pointed out by \citet{Mas10}, there is a continuous 
link that connects \emph{some} CEMP-no stars with CEMP-s stars,
while the carbon enhancement of other CEMP-no stars
may have a different origin.

Models of core-collapse supernova yields from 
Population~III (or near metal-free) stars 
do entertain high C yields relative to Fe (e.g. \citealt{WooWea95}). 
Population III enrichment is a particularly intriguing explanation 
for the origin of the carbon enhancement in some CEMP stars, 
since the three most Fe-poor halo stars all exhibit carbon
enhancements relative to iron \citep{Chr02,Fre05,Nor07}. 
Moreover, the fraction of metal-poor stars that 
exhibit a carbon-enhancement \emph{increases} 
with decreasing metallicity \citep{BeeChr05}. 
In this Population III enrichment scenario, the carbon enhancement 
in the extremely metal-poor regime reflects the initial composition
of the gas from which these stars formed, 
rather than resulting from mass transfer 
from an evolved companion.

Whilst the physics behind core-collapse supernovae is poorly constrained, 
several parameterised models have been developed to calculate the expected 
yields from zero-metallicity progenitors to compare with the observations 
of the most Fe-poor CEMP stars, as well as the more `normal' non-CEMP stars. 
The most important (and largely unknown) quantities are the degree of 
mixing during the supernova explosion and the amount of fallback onto the 
remnant black hole thereafter \citep{UmeNom03}. \citet{UmeNom03} 
parameterised both quantaties, then suitably selected the appropriate values 
that reproduce the observed stellar abundance patterns. \citet{HegWoo08}, on 
the other hand, parameterise only the mixing parameter by applying a running 
boxcar filter over the star following the explosion. This prescription 
successfully reproduces the chemical composition of 
the extremely metal-poor (non-CEMP) halo stars from the study
by \citet{Cay04}, as well as that of the most Fe-poor 
CEMP stars HE0107$-$5240 \citep{Chr02} and 
HE1327$-$2326 \citep{Fre05}. 
 \citet{JogWooHeg09} extended this work by mapping 
the one-dimensional models by \citet{HegWoo08} onto a two-dimensional 
grid to follow Rayleigh-Taylor induced mixing after explosive nuclear burning. 
Whilst the \citet{JogWooHeg09} models are physically motivated, they are unable 
to reproduce the relatively high levels of nitrogen enrichment that are 
observed in the most Fe-poor stars, nor are they able to produce sufficient 
Fe-peak elements.

These concerns are alleviated with models that include rotation 
(e.g. \citealt{MeyEksMae06,Hir07,Mey10}), since rotation induces additional
mixing and mass loss in the pre-supernova phase of low-metallicity 
stars. The only simulation available to date that investigates the 
effects of both rotational and Rayleigh-Taylor mixing, as well as 
incorporating a realistic prescription of fallback, are those presented 
recently by \citet{Jog10a} (see also \citealt{Jog10b}). By introducing 
rotation in their zero-metallicity models, \citet{Jog10a} report an 
increased nitrogen yield, as well as an increased Fe-peak element yield,
however, these models are not able to reproduce the high CNO to Fe 
ratios observed in the most metal-poor stars. 
In summary, it is still a matter of some debate whether  
these extremely metal-poor early Population~II stars: (i) were borne 
out of gas which had previously been carbon-enriched by the first stars;
or (ii) received 
a CNO top-up from a companion star;  and/or (iii) exhibit 
a degree of self-pollution from 
their own nucleosynthesis.

The first possibility, that at least some CEMP stars were borne out 
of carbon-enhanced gas,  is given
additional support from the discovery
reported here of an extremely metal-poor DLA at $z_{\rm abs} = 2.3400972$
exhibiting a carbon enhancement akin to that 
measured in Galactic CEMP stars. 
No other case was known until now, given (i) the rarity of
DLAs with [Fe/H]\,$< -2$,
which lie in the tail of the metallicity distribution of DLAs
\citep[e.g.][]{Pon08}, and (ii) the difficulty in measuring
[C/H] in DLAs, where the relevant absorption lines are often
saturated even in the VMP regime.
We speculate that this DLA may be the `missing link' between 
the first few generations of stars and some of the CEMP stars
in the Galactic halo.

The paper is organized as follows.
Section~2 summarizes the  observations and data reduction. 
We analyse the absorption lines in the damped \Lya\ system
and deduce corresponding element abundances in Section~3. 
In Section 4 we scrutinize potential issues that could 
masquerade as a carbon enhancement.
After ruling out these possibilities, we
discuss in Section~5 possible origins of this enhancement, 
and consider the implications of our finding for models of 
CEMP stars. Finally, we summarize our results and draw our conclusions 
in Section~6.

%%%%%%%%%%%%%%%%%%%%%%%%%%%%%%%%%%%%%%%%%
\section{Observations and Data Reduction}
\label{sec:obs}
%%%%%%%%%%%%%%%%%%%%%%%%%%%%%%%%%%%%%%%%%

The $m_{\rm r} = 18.89$, $z_{\rm em} = 2.42$ QSO J0035$-$0918 
was selected for our on-going survey for VMP DLAs \citep{Coo10b} 
on the basis of its SDSS spectrum which 
shows a DLA at $z_{\rm abs}  \simeq 2.340$ with no apparent associated 
metal lines. Such cases are highly suggestive of narrow
and weak metal absorption lines which are undetectable at the coarse resolution
and signal-to-noise ratio (S/N) of the discovery SDSS spectra.

Follow-up observations of J0035$-$0918 were made with the
Magellan Echellette (MagE) spectrograph \citep{Mar08}
on the Magellan~\textsc{ii} Clay telescope on the nights of
2008 December 30 and 31 in good conditions with 1\,arcsec seeing.
We took $2\times 2400$\,s exposures using the
standard setup with 1$\times$1 binning and a 1.0\,arcsec slit
giving a resolving power  $R \equiv \lambda/\Delta \lambda \simeq 4100$. 
Standard calibrations were taken following the recommendations by
Simcoe, Hennawi \&\ 
Williams\footnote{see \url{http://web.mit.edu/rsimcoe/www/MagE/mage.html}}.
The data were reduced using a custom set of IDL routines
written by G.~D.~Becker and described in \citet{Bec06}.  
The MagE spectrum confirmed the VMP DLA nature of the 
$z_{\rm abs}=2.3400972$ absorption system, by showing clear
damping wings to the \Lya\ absorption line (see top panel
of Figure~\ref{fig:DLA}) and unusually
weak associated metal lines. 

Encouraged by these initial
indications, we subsequently 
observed J0035$-$0918 on the night of 2009 December 9
with the High Resolution Echelle Spectrograph \citep{Vog94} 
on the Keck~\textsc{i} telescope
under good conditions with sub-arcsecond seeing, 
for a total integration time of 16200\,s,
divided into 6 exposures of 2700\,s, 
resulting in a signal-to-noise ratio near 4500\,\AA\ 
of ${\rm S/N} \simeq 18$.
We used the $1.148$\,arcsec wide slit which, when uniformly illuminated,
provides a nominal spectral resolution 
$R \equiv \lambda/\Delta \lambda = 37\,000$.
From our spectra, we measure $R \simeq 41\,000$ which
corresponds to a velocity 
full-width at half maximum 
${\rm FWHM} = 7.3\,\,{\rm km\,\,s}^{-1}$,
sampled by $\sim3$ pixels.
We employed the UV cross-disperser 
which covers the wavelength range 3100--6000\,\AA\
with $\sim 70$\,\AA-wide gaps near 4000\,\AA\ and $5000$\,\AA.

The HIRES spectra were reduced with 
the \textsc{makee} data reduction pipeline developed by Tom Barlow,
which performs the usual steps of flat-fielding, order tracing, background
subtraction, and extraction of the final 1-D spectrum. 
The data were wavelength calibrated
by reference to the spectrum of a ThAr lamp, 
and mapped onto a vacuum heliocentric wavelength scale.
The extracted spectra were merged and then normalized
by dividing out the QSO continuum and emission lines, using the
software package \textsc{uves popler}, maintained by 
Michael Murphy\footnote{\textsc{uves popler} is available from\\
http://astronomy.swin.edu.au/$\sim$mmurphy/UVES\_popler}.
Following this step, all available metal absorption lines associated with the DLA
were extracted in $\pm 150$\,km~s$^{-1}$ windows around the pixel with highest optical depth.
A further fine correction to the continuum was then applied if necessary.

%%%%%%%%%%%%%%%%%%%%%%%%%%%%%%%%%%%%%%%%%%%%%%%%
\section{Profile Fitting and Abundance Analysis}
\label{sec:DLA}
%%%%%%%%%%%%%%%%%%%%%%%%%%%%%%%%%%%%%%%%%%%%%%%%

%%%%%%%%%%%%%%%%%%
%%%  Figure 1  %%%
%%%%%%%%%%%%%%%%%%
\begin{figure*}
  \centering
  \includegraphics[angle=0,width=160mm]{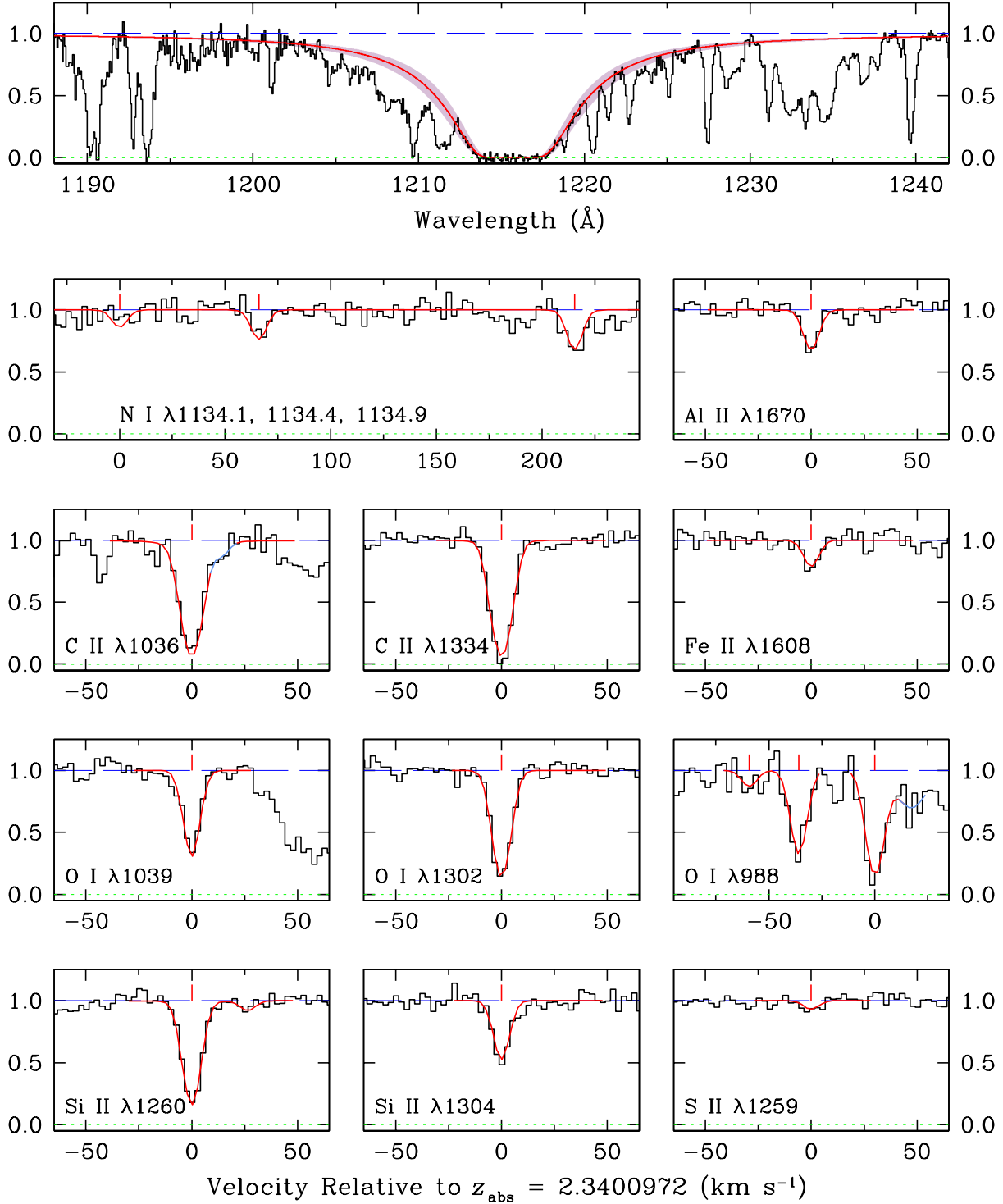}
  \caption{ 
Selected absorption lines in the $z_{\rm abs}=2.3400972$ DLA towards
J0035$-$0918. The data are shown with black histograms, while the red
continuous lines are model fits to the line profiles.
\textit{Top panel:} Portion of the MagE spectrum of J0035$-$0918 
encompassing the damped \Lya\ line, together with the
theoretical Voigt profile (red line) for a neutral
hydrogen column density $\log[N({\rm H}\,\textsc{i})/{\rm cm}^{-2}] = 20.55 \pm 0.1$.
The remaining panels display portions of the HIRES spectrum
near metal lines of interest, together with model profiles
generated with \textsc{vpfit} as described in Section~\ref{sec:col_dens}.
This DLA consists of a single absorption component at $z_{\rm abs}=2.3400972$
with a small velocity dispersion, $b = \sqrt{2} \sigma = 2.4$\,km~s$^{-1}$.
The weak absorption feature centred at $+26\,{\rm km~s}^{-1}$ 
in the \SiII\,$\lambda 1260.4221$ panel is probably 
\FeII\,$\lambda 1260.533$ absorption in the DLA, although its strength
is below our $3 \sigma$ detection limit. The red wings of both
C\,{\sc ii}\,$\lambda 1036$ and \OI$\,\lambda988$ are
blended with a weak \Lya\ forest  line
 indicated by a continuous blue line.
In all plots the $y$-axis scale is residual intensity.
  }
  \label{fig:DLA}
\end{figure*}

Table~\ref{tab:lines} lists all the metal absorption lines
in the $z_{\rm abs}=2.3400972$ DLA detected 
in our HIRES spectrum of J0035$-$0918, together with a few interesting
transitions which are below the detection limit of the data.
For each line, we give the rest-frame wavelength and 
oscillator strength that we adopted for this work, 
as well as the measured equivalent width and associated 
random and systematic errors. The latter
were determined by
shifting the continuum by $\pm\sigma/\sqrt{n}$
(where $\sigma$ is the $1\sigma$ error spectrum, 
and $n$ is the number of independent resolution elements over
which the equivalent width integration was carried out),
%i.e. $n=$ number of absorption pixels / sampling size),
and recalculating the equivalent width.
For the undetected transitions, we quote the 
$3\sigma$ limiting rest-frame
equivalent width, using as a guide the profile 
of the weakest absorption line, \FeII~$\lambda 1608$,
which we detect at the $5 \sigma$ confidence limit
(see Table~\ref{tab:lines}). Examples of metal absorption
lines are reproduced in Fig.~\ref{fig:DLA}.

\begin{table}
%\centering
    \caption{\textsc{Metal lines in the $z_{\rm abs} = 2.3400972$ DLA towards J0035$-$0918}}
    \begin{tabular}{lllccc}
    \hline
   \multicolumn{1}{l}{Ion}
& \multicolumn{1}{l}{Wavelength$^{\rm a}$}
& \multicolumn{1}{c}{$f^{\rm a}$}
& \multicolumn{1}{c}{$W_0^{\rm b}$}
& \multicolumn{1}{c}{$\delta W_0^{\rm c}$}
& \multicolumn{1}{c}{$\delta W_{\rm 0,cont}^{\rm d}$}\\
    \multicolumn{1}{c}{}
& \multicolumn{1}{c}{(\AA)}
&  \multicolumn{1}{c}{}
& \multicolumn{1}{c}{(m\AA)}
& \multicolumn{1}{c}{(m\AA)}
& \multicolumn{1}{c}{(m\AA)} \\
    \hline
\CII    & 1036.3367      & 0.118     & $39$             & $2$   &  $1$   \\
\CII    & 1334.5323      & 0.1278    & $54$             & $2$   &  $1$   \\
\NI     & 1134.1653      & 0.0146    & $< 5.3^{\rm e}$  & \dots & \dots  \\
\NI     & 1134.4149      & 0.0278    & $6.2$            & $1.7$ &  $0.5$ \\
\NI     & 1134.9803      & 0.0416    & $12.5$           & $1.6$ &  $0.5$ \\
\NII    & 1083.9937      & 0.111     & $< 4.6^{\rm e}$  & \dots & \dots  \\
\OI     &  971.7382      & 0.0116    & $24$             & $3$   &  $1$   \\
\OI     &  988.5778      & 0.000553  & $< 6.0^{\rm e}$  & \dots & \dots  \\
\OI     &  988.6549      & 0.0083    & $23$             & $2$   &  $1$   \\
\OI     &  988.7734      & 0.0465    & $38$             & $3$   &  $1$   \\
\OI     & 1039.2304      & 0.00907   & $23$             & $2$   &  $1$   \\
\OI     & 1302.1685      & 0.048     & $42$             & $2$   &  $1$   \\
\AlII   & 1670.7886      & 1.740     & $15$             & $2$   &  $0.5$ \\
\SiII   &  989.8731      & 0.171     & $15$             & $2$   &  $1$   \\
\SiII   & 1193.2897      & 0.582     & $28$             & $3$   &  $2$   \\
\SiII   & 1260.4221      & 1.18      & $39$             & $2$   &  $0.5$ \\
\SiII   & 1304.3702      & 0.0863    & $21$             & $2$   &  $1$   \\
\SiII   & 1526.7070      & 0.133     & $34$             & $2$   &  $1$   \\
\SII    & 1253.805       & 0.0109    & $< 2.0^{\rm e}$  & \dots & \dots  \\
\SII    & 1259.5180      & 0.0166    & $3.7$            & $0.8$ &  $0.3$ \\
\FeII   & 1260.533       & 0.0240    & $< 2.5^{\rm e}$  & \dots & \dots  \\
\FeII   & 1608.4509      & 0.0577    & $10$             & $2$   &  $0.5$ \\
\hline
\end{tabular}
%\smallskip
\begin{flushleft}
$^{\rm a}$ Laboratory wavelengths and $f$-values from \citet{Mor03}.\\
%~~\,with updates by \citet{Jen06}.\\
$^{\rm b}$ Rest-frame equivalent width.\\
$^{\rm c}$ Random error on the equivalent width $W_0$.\\
$^{\rm d}$ Systematic error on $W_0$ from $1 \sigma$ uncertainty in the continuum placement.\\
$^{\rm e}$ $3\sigma$ upper limit on the rest-frame equivalent width.\\
\end{flushleft}
\label{tab:lines}
\end{table}

As can be appreciated from inspection of Table~\ref{tab:lines}
and Fig.~\ref{fig:DLA}, the metal lines in this DLA are very weak
and narrow, with equivalent widths $W_0 < 55$\,m\AA\ (the strongest
line being C\,{\sc ii}\,$\lambda 1334$), and with the absorption
taking place in a single velocity component with FWHM\,$ < 10$\,km~s$^{-1}$.
The weakness of the absorption limits our detection to the 
intrinsically most abundant elements of the periodic table, 
C, N, O, Al, Si, and Fe; on the other hand, the wide wavelength coverage of
the echelle spectra, which reach well into the far-UV, gives access to several
transitions of differing $f$-values for most of these elements.

\subsection{Column Densities}
\label{sec:col_dens}

We begin our abundance analysis by measuring the column density of
neutral hydrogen. To this end, we used the MagE spectrum of the 
QSO, the relevant portion of which is reproduced in the top panel
of Fig.~\ref{fig:DLA}, because at $z_{\rm abs}=2.3400972$
the damped \Lya\ line falls on a gap between two of the CCDs
in the HIRES detector mosaic. Even at the coarser resolution of MagE
(compared to HIRES), the broad damped \Lya\ line is fully resolved
and no loss of accuracy results in the derivation of $N$(H\,{\sc i}).
Fitting a Voigt profile to the line, we deduced 
$\log [N$(\HI)/cm$^{-2}] = 20.55 \pm 0.10$; the corresponding
theoretical Voigt profile is overlaid on the MagE spectrum
in the top panel of Fig.~\ref{fig:DLA}.

In the next step, we determined the 
Doppler parameter of the absorbing gas, $b$ (km~s$^{-1}$), 
and the 
column density of the metal ions, $N$(X) (cm$^{-2}$),
by fitting the corresponding line profiles with 
\textsc{vpfit}\footnote{\textsc{vpfit} is available 
from http://www.ast.cam.ac.uk/${\sim}$rfc/vpfit.html}, 
which simultaneously fits multiple Voigt profiles to several atomic transitions,
returning the values of $N$(X) and $b$, together with
the associated errors, that minimize the $\chi^{2}$ between the
data and the model.
We tied the redshift and the Doppler parameter to be the same
for all of the absorption lines listed in Table~\ref{tab:lines}, 
which is justified if the neutrals and first ions are
kinematically associated with the same gas 
(we relax this assumption later). 
With these constraints, \textsc{vpfit} converged 
to a best-fitting model consisting of a 
single absorption component with redshift
$z_{\rm} = 2.3400972 \pm 0.0000008$ 
and Doppler parameter $b = 2.36 \pm 0.08$ km~s$^{-1}$.
The corresponding column densities 
are listed in Table~\ref{tab:cd}.
Examples of 
the theoretical line profiles 
generated by \textsc{vpfit} are shown superimposed
on the data in Fig.~\ref{fig:DLA}.

The weakest feature in our data is \SII\,$\lambda 1259$.
When this line is included in the \textsc{vpfit}
fitting procedure (see
bottom right panel of Fig.~\ref{fig:DLA}), 
we derive a column density 
$\log N$(\SII)/${\rm cm}^{-2} = 13.08 \pm\ 0.10$.
However, since this absorption line is only
significant at the $\sim 4.5 \sigma$ level,
we conservatively consider the above value to be 
an upper limit to the column density of S\,{\sc ii}.

With the usual assumption that the ions observed
are the dominant stage of the corresponding elements in H\,{\sc i}
gas, so that corrections for unseen ion stages and/or
the presence of ionized gas are negligible (we review this assumption
in Section~\ref{sec:ion_corr} below),
it is straightforward to deduce the abundances of the elements
concerned by simply dividing the column densities in
Table~\ref{tab:cd} by $N$(H\,{\sc i}).
Comparison with the solar abundance scale of \citet{Asp09}
then leads to the abundance pattern listed 
in Table~\ref{tab:abund} and illustrated graphically
in Fig.~\ref{fig:abund_plot}.

\begin{table}
\centering
    \caption{\textsc{Ion column densities in the $z_{\rm} = 2.3400972$
    DLA towards J0035$-$0918}}
    \begin{tabular}{@{}lc}
    \hline
  \multicolumn{1}{l}{Ion}
& \multicolumn{1}{c}{$\log N$(X)/${\rm cm}^{-2}$}\\
    \hline
\HI    &  20.55 $\pm$ 0.10  \\
\CII   &  15.47 $\pm$ 0.15  \\
\NI    &  13.51 $\pm$ 0.06  \\
\OI    &  14.96 $\pm$ 0.08  \\
\AlII  &  11.73 $\pm$ 0.05  \\
\SiII  &  13.41 $\pm$ 0.04  \\
\SII   &  $\le 13.08$       \\
\FeII  &  12.98 $\pm$ 0.07  \\
    \hline
    \end{tabular}
   \smallskip
   \label{tab:cd}
\end{table}

\subsection{Ionization  Corrections}
\label{sec:ion_corr}

In DLAs, it is usually assumed that the metals within the absorbing \HI\ 
gas reside in a single dominant ionization stage, X\,\textsc{n}, so that the
total abundance of an element is given by
\begin{equation}\label{eq:ics}
%\log \bigg[\frac{N(\mathrm{X})}{N(\mathrm{H})}\bigg] = \mathrm{IC(X)} + 
%\log \bigg[\frac{N(\mathrm{X}\,\textsc{n})}{N(\mathrm{H}\,\textsc{i})}\bigg]
[{\rm X/H}] = [{\rm X}\,\textsc{n}/\HI] + \mathrm{IC(X)}
\end{equation}
where the ionization correction, IC(X), is typically negligible.
In general, it is safe to assume IC(X)\,$ \simeq  0.0$ for gas with high \NHI,
since the gas is self-shielded from
ionizing radiation \citep[e.g.][]{Vla01}.
Of course, if the gas does not reside in a single dominant ionization stage,
or some amount of X\,\textsc{n} is associated with \HII\ gas,
we may respectively under- or over-estimate the abundance
of a given element [positive or negative IC(X)].

\begin{figure}
 \vspace{1cm}
  \centering
  \includegraphics[angle=0,width=80mm]{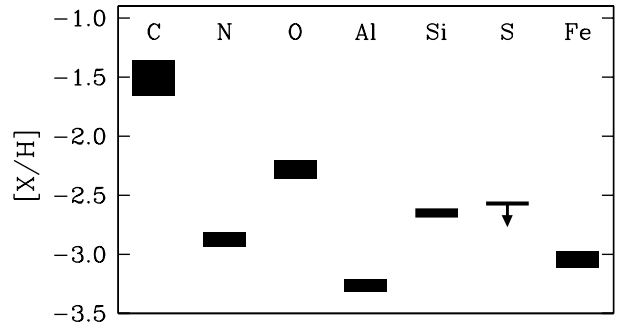}
  \caption{ 
Element abundances in the $z_{\rm abs} =  2.3400972$
DLA towards J0035$-$0918 compared to the solar values as compiled by
\citet{Asp09}. The height of each box represents the 
uncertainty in each element abundance; the 
$3\sigma$ upper limit for S is indicated by the bar and arrow.
  }
  \label{fig:abund_plot}
\end{figure}

To gauge the extent of such corrections, 
we used the
\textsc{cloudy} photoionization software \citep{Fer98}
to model the DLA
as a slab of gas with uniform density in the range
$-3 < \log[n({\rm H})/{\rm cm}^{-3}] < 3$,
exposed to the \citet{HarMad01} metagalactic
ionizing background and the cosmic microwave background,
both at the redshift of the DLA.
Adopting the solar abundance scale of \citet{Asp09},
we globally scaled the metals to log $Z_{\rm DLA}\,/\,Z_{\odot} = -2.75$
(an approximate average metallicity---see Fig.~\ref{fig:abund_plot}).
Once the column density of the DLA was reached, we stopped the
simulations and output the resulting ion column densities.
Using these computed column densities, we are then able to derive
the ionization corrections from Eq.~\ref{eq:ics}.

\begin{table}
\centering
    \caption{\textsc{Element abundances in the $z_{\rm} = 2.3400972$
    DLA towards J0035$-$0918}}
    \begin{tabular}{@{}cccc}
    \hline
  \multicolumn{1}{c}{Element}
& \multicolumn{1}{c}{$\log \epsilon{\rm (X)}_{\rm DLA}^{\rm a}$}
& \multicolumn{1}{c}{$\log \epsilon{\rm (X)}_{\odot}^{\rm a,b}$}
& \multicolumn{1}{c}{[X/H]$_{\rm DLA}^{\rm c}$}\\
    \hline
C   &  $6.92$   &  $8.43$  &  $-1.51$  \\
N   &  $4.96$   &  $7.83$  &  $-2.87$  \\
O   &  $6.41$   &  $8.69$  &  $-2.28$  \\
Al  &  $3.18$   &  $6.44$  &  $-3.26$  \\
Si  &  $4.86$   &  $7.51$  &  $-2.65$  \\
S   & $\le 4.53$ &  $7.14$  &  $\le -2.61$ \\
Fe  &  $4.43$   &  $7.47$  &  $-3.04$  \\
    \hline
    \end{tabular}
    %\smallskip
\newline
\begin{flushleft}
$^{\rm a}${$\log \epsilon{\rm (X)}=12+\log N({\rm X})/N({\rm H})$}.
\\
$^{\rm b}${Asplund et al. (2009)}.
\\
$^{\rm c}${[X/H]$_{\rm DLA} \equiv \log \epsilon{\rm (X)}_{\rm DLA} - \log \epsilon{\rm (X)}_{\odot}$},
with errors as listed in Table~2.\\
\end{flushleft}
   \label{tab:abund}
\end{table}

\begin{figure*}
  \centering
 {\hspace{-0.25cm} \includegraphics[angle=0,width=80mm]{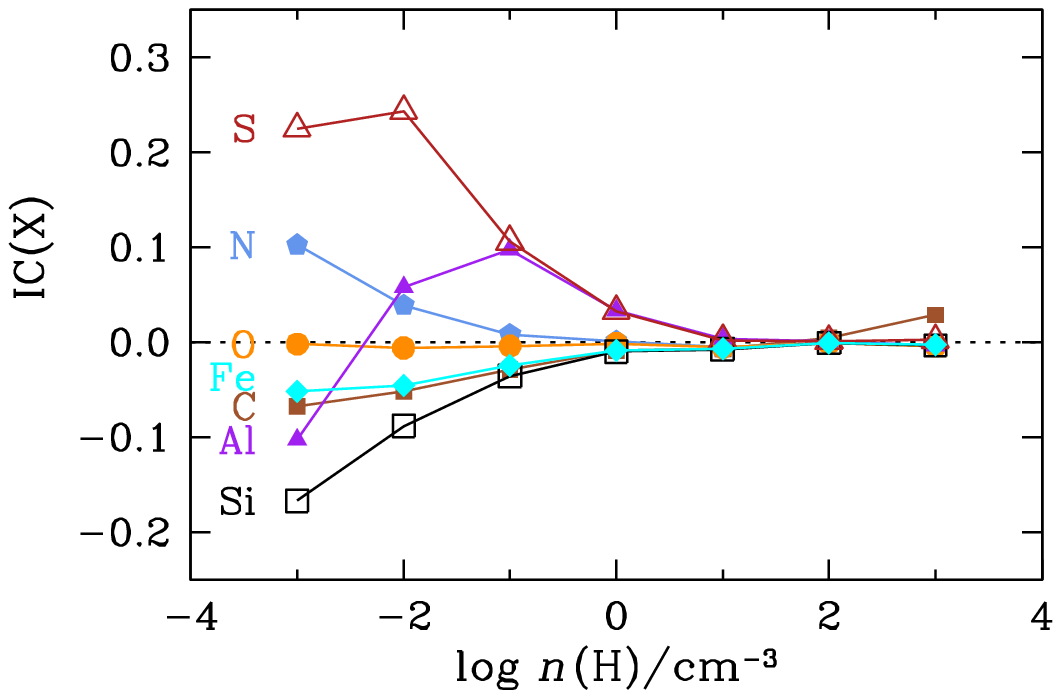}}
 {\hspace{1.05cm} \includegraphics[angle=0,width=80mm]{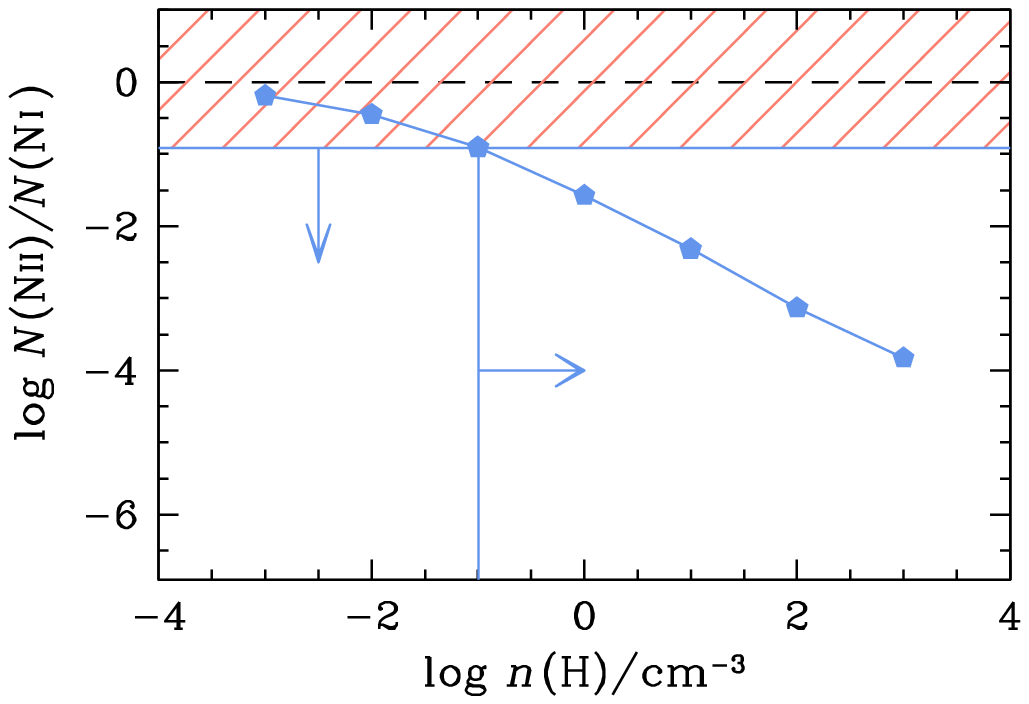}}
  \caption{
\emph{Left panel}: 
Ionization corrections, as defined by Eq.~\ref{eq:ics}, for 
elements of interest (symbols connected by a solid line), 
as a function of the volume density of the gas in the
$z_{\rm abs} =  2.3400972$ DLA.
\emph{Right panel}: Ratio of the computed column densities for the successive
ion stages of nitrogen (symbols connected by a solid line), plotted as a function
of the volume density.
The solid horizontal line indicates the upper limit we measure for
$\log N{\rm (N\,\textsc{ii})}$/$N{\rm (N\,\textsc{i}})$,
whilst the vertical line with an arrow indicates the corresponding lower limit on the
volume density of the gas, $\log [n{\rm (H)/cm}^{-3}] \geq -1.0$.
Both plots were produced with \textsc{cloudy} photoionization
calculations; see the text for further details.
  }
  \label{fig:ics_gd}
\end{figure*}

The results of our \textsc{cloudy} simulations are shown in
Fig.~\ref{fig:ics_gd}. The ionization corrections
appropriate to the DLA under investigation depend on
the volume density of the gas, which can be inferred
from the ratio of successive ion stages 
(see right panel of Fig.~\ref{fig:ics_gd}).
The only element for which we have this information is nitrogen
and, even then, we can only derive a $3 \sigma$ upper limit
to the \NII\ column density from the $3\sigma$ upper limit on the
rest frame equivalent width 
of the undetected N\,{\sc ii}\,$\lambda 1084$ line
(see Table~\ref{tab:lines}),
leading to the upper limit $\log N{\rm (N\,\textsc{ii})/}N{\rm (N\,\textsc{i}})\,\leq -0.91$.
This affords a lower limit on the volume density 
of $\log [n{\rm (H)/cm}^{-3}] \geq -1.0 $
(see right-hand panel of Fig.~\ref{fig:ics_gd}).
Referring now to the left-hand panel of Fig.~\ref{fig:ics_gd},
it can then be seen that, when the gas density is greater
than 0.1\,cm$^{-3}$, the 
ionization corrections for the ions of interest here
are indeed small, ${\rm IC} \simlt 0.1\,{\rm dex}$.
These values are comparable to the uncertainties in the 
ion column densities (see Table~\ref{tab:cd}) which
justifies our assumption that ${\rm [X/H]} \sim$ [X\,\textsc{n}/\HI],
and that ionization corrections are not a serious concern.

It is worth noting, in passing, that even in the absence of 
several density-sensitive ion ratios, low values of $n$(H) 
are in general unlikely in these metal-poor DLAs with such simple 
velocity structure. The column density $\log [N(\HI)/{\rm cm}^{-2}] = 20.55$
implies a linear size $l > 1$\,kpc along the line of sight 
if  $\log [n{\rm (H)/cm}^{-3}] <  -1$.
It seems unlikely that the structures giving rise to DLAs
with properties similar to that considered here could
have such large physical extent 
while maintaining very quiescent
kinematics, with internal velocity dispersions
of only a few km~s$^{-1}$.

%%%%%%%%%%%%%%%%%%
%%%  Figure 2  %%%
%%%%%%%%%%%%%%%%%%

\section{A masquerading carbon enhancement?}

Returning to Fig.~\ref{fig:abund_plot}, we note that
with [O/H]\,$ = -2.28$ and [Fe/H]\,$ = - 3.04$ the
$z_{\rm} = 2.3400972$ DLA in line to
J0035$-$0918 is among the most metal-poor known
\citep{Pet08,Pen10}. 
However, the most striking feature of the abundance 
pattern in Fig.~\ref{fig:abund_plot} is the 
overabundance of carbon relative to all other elements.
Thus, for example, [C/O]\,$ = +0.77$, which implies 
an overabundance
of carbon relative to oxygen by a factor of $\sim 6$;
similarly, [C/Fe]\,$ = +1.53$, or $\sim 35$ times solar!
Given such extreme values, it is important to consider
what factors, if any, may have led us to spuriously overestimate
the abundance of carbon. To this end, we performed several
tests, which we now discuss. The upshot of these tests 
is that we could not find any plausible reason why
the abundance of carbon would have been significantly
overestimated. Thus, readers who are less interested in
the details of these checks may skip the rest of this section
and move straight to the discussion in Section~\ref{sec:discuss}.

\subsection{Incorrect background subtraction? }

%%%%%%%%%%%%%%%%%%
%%%  Figure 3  %%%
%%%%%%%%%%%%%%%%%%

First we considered the possibility that the background
may have been oversubtracted in the proximity of the 
C\,{\sc ii} absorption lines; if the zero level had been incorrectly
determined, this may lead us to overestimate the apparent
optical depth of the absorption.
While we cannot categorically rule out this possibility,
as we do not have any independent measures of the 
zero level (such as saturated absorption lines with flat cores)
in the immediate vicinity of the 
C\,{\sc ii} lines, we note the following:
(i) there are a number of strongly saturated 
\Lya\ absorption lines in the \Lya\ forest
(shortwards of $\lambda_{\rm obs} = 4158$\,\AA) and
none of them show a systematic offset
of their cores from zero residual intensity;
(ii) the apparent optical depths of the \emph{two}
C\,{\sc ii} transitions covered by our HIRES spectrum,
$\lambda 1334$ and $\lambda 1036$, are mutually consistent
(see Fig.~\ref{fig:DLA}); 
thus, if the background level were incorrect,
it would have to have
been oversubtracted by the same fractional amount of the QSO
continuum at these two wavelengths, separated by 995\,\AA\ 
in the observed spectrum (at $z_{\rm abs} = 2.3400972$).
We also inspected the raw 2-D HIRES frames in the region
of the two best observed C\,{\sc ii} and O\,{\sc i} lines,
$\lambda 1334$ and $\lambda 1302$ and found that they
both fall close to the peak of the echelle blaze function,
near the centre of the HIRES detector, where 
the data are of the highest S/N.
We conclude that 
the recorded excess of C is unlikely
to be an artifact of the data reduction process.

\subsection{Profile fitting 1: gas kinematics}
\label{sec:gas_kinem}

Next we looked critically at the profile fitting procedure.
Our \textsc{vpfit} modelling described in Section~\ref{sec:col_dens}
assumed that the neutrals
and singly ionized species arise from gas with the same Doppler parameter. 
We relaxed this constraint by fitting separately the absorption lines
of the first ions (two C\,{\sc ii}, five Si\,{\sc ii} lines and one line each
of Al\,{\sc ii} and Fe\,{\sc ii}) and the neutrals (five O\,{\sc i} and
two N\,{\sc i} lines).
In this case, \textsc{vpfit} converged to best fitting
Doppler parameters of $b = 2.30 \pm 0.09$
for the first ions and $b = 2.6\pm 0.2$  for the neutrals.
With these parameters the \CII\ column density is \emph{higher}
by 0.1\,dex than the value listed in Table~\ref{tab:cd} 
(that is, the value obtained by tying $b$ to be the same
for all species), while $N$(O\,{\sc i}) is \emph{lower} by 0.1\,dex. 
The column densities of other elements remain essentially unchanged.
In other words, kinematically decoupling neutrals and first ions
has the net effect of further \emph{increasing} the [C/O] overabundance
relative to the values in Table~\ref{tab:abund} and Fig.~\ref{fig:abund_plot}.
Note also that the above values of $b$ are still consistent within the 
errors with the best-fitting $b = 2.36 \pm 0.08$ deduced in
Section~\ref{sec:col_dens} by tying all the absorption lines together.
Finally, fitting \emph{only} the \CII\ lines, without reference
to any other transition, yielded 
$\log [N{\rm (\CII)/cm}^{-2}] =15.42 \pm 0.59$, which is only 
0.05\,dex lower than the value listed in Table~\ref{tab:cd},
albeit with a larger error
(but with much the same
redshift and Doppler parameter:
$z_{\rm abs} = 2.340096 \pm 0.000002$ and 
$b = 2.4 \pm 0.3$\,km~s$^{-1}$).

%%%%%%%%%%%%%%%%%%
%%%  Figure 4  %%%
%%%%%%%%%%%%%%%%%%
\begin{figure*}
  \centering
  \includegraphics[angle=0,width=115mm]{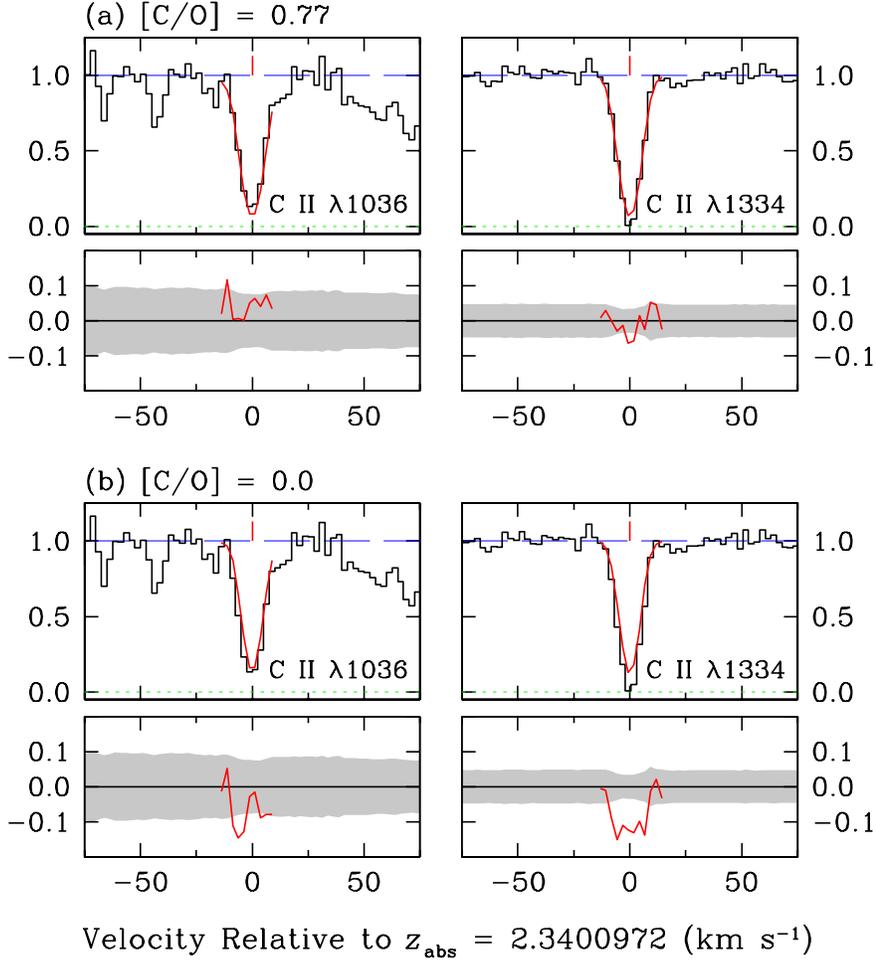}
  \caption{ 
Comparison between observed C\,{\sc ii} line profiles (black histograms)
and theoretical profiles computed with \textsc{vpfit} (red continuous lines).
Below each plot we also show the difference between computed and observed
residual intensities (red lines) compared with the $1\sigma$ error spectrum
of the data (grey area). \textit{(a):} best fitting model as described in 
Section~\ref{sec:col_dens} with the $\log [N({\rm \CII})/{\rm cm}^{-2}] = 15.47$
as in Table~\ref{tab:cd}.
\textit{(b):} best fitting model obtained by forcing
$\log [N({\rm \CII})/{\rm cm}^{-2}] = 14.70$ so that
[C/O]\,= \,0.0 (see text
for further details).
This model provides a poorer 
fit to both \CII\ absorption lines, 
as demonstrated by the mismatch between the model
subtracted from the data and the $1\sigma$ error spectrum
(bottom panels).
}
  \label{fig:co_solar}
\end{figure*}

\subsection{Profile fitting 2: instrumental resolution}

In order to compare theoretical and observed absorption line
profiles, \textsc{vpfit} requires knowledge of the instrumental
broadening function, normally assumed to be a Gaussian with
FWHM corresponding to the spectral resolution. The nominal
resolution of spectra recorded through the $1.148$\,arcsec wide 
entrance slit of HIRES employed in our observations is
FWHM\,$ = 8.1$\,km~s$^{-1}$ (see http://www2.keck.hawaii.edu/inst/hires/),
but this applies to a uniformly illuminated slit.
Since the seeing was consistently better than 
$1.148$\,arcsec during our observations of J0035$-$0918,
it is likely that the actual resolution of our data is somewhat
better than the nominal value.
The narrowest features in the spectrum of J0035$-$0918
are the metal absorption lines of the $z_{\rm abs} = 2.3400972$ DLA
themselves.
We therefore estimated the true spectral resolution---measured
by the Doppler parameter $b_{\rm instr} \equiv 0.6006$\,FWHM---by 
varying $b_{\rm instr}$ in small steps from 5.0 to 3.7\,km~s$^{-1}$
(i.e. between FWHM\,$= 8.3$ and 6.2\,km~s$^{-1}$),
fitting all of the absorption lines as described in Section~\ref{sec:col_dens},
and minimizing the $\chi^2$ between the model and observed
profiles as a function of $b_{\rm instr}$. This procedure gave a 
well-defined $b_{\rm instr,min} = 4.4$\,km~s$^{-1}$, corresponding
to FWHM\,=\,7.3\,km~s$^{-1}$, or $R = 41\,000$, which is
the value used in all the profile fitting described in Sections~\ref{sec:col_dens}
and \ref{sec:gas_kinem}.

It is obviously important to test the sensitivity of our results
to the value of $b_{\rm instr}$ adopted. Within the range of 
values tested, $b_{\rm instr} = 5.0$--3.7\,km~s$^{-1}$, the 
column density of \CII\ and the corresponding
element ratios changed by no more
than $\pm 0.1$\,dex compared to the values in  
Table~\ref{tab:cd} which refer to the case 
$b_{\rm instr,min} = 4.4$\,km~s$^{-1}$.
%For reference, the lowest value tested, $b_{\rm instr} = 3.7$\,km~s$^{-1}$, 
%corresponds to the spectral resolution $R = 48\,000$ delivered 
%by the  0.861\,arcsec wide slit of HIRES.
On the basis of the tests carried out in this and the preceding subsection, 
we conclude that the overabundance of carbon is not an artifact of the 
profile fitting analysis of the absorption lines.

\subsection{Monte Carlo simulations}

The $z_{\rm} = 2.3400972$ DLA in 
J0035$-$0918 is unique so far in showing such a marked
overabundance of carbon. More typically, DLAs with
${\rm [O/H]} < -2$  have  [C/O] $\simeq -0.2$ and
at most [C/O] $\simeq 0.0$ 
[i.e. (C/O) = (C/O)$_{\odot}$; \citealt{Pet08, Coo10b}].
It is therefore worthwhile considering to what extent
a solar C/O ratio is excluded by our data.
We illustrate this test in Fig.~\ref{fig:co_solar}, where 
we compare the line profiles for the two C\,{\sc ii}
absorption lines generated with: \textit{(upper panels)}
the best-fitting model returned by \textsc{vpfit},
and \textit{(lower panels)} a model in which 
$N$(C\,{\sc ii}) has artificially been fixed 
at the column density corresponding
to [C/O]\,=\,0.0 (that is $\log [N({\rm \CII})/{\rm cm}^{-2}] = 14.70$
as opposed to 15.47 as in Table~\ref{tab:cd}).
While the former results in a $\chi^{2}/{\rm dof} = 13.9/22$,
the latter fits are worse, with $\chi^{2}/{\rm dof} = 86.6/22$.
The higher value of $\chi^{2}$ is reflected by the
higher residuals, both in the core and the wings of the
C\,{\sc ii}\,$\lambda 1334$ line in particular,
as can be appreciated from  close examination of
Fig.~\ref{fig:co_solar}. The noisier C\,{\sc ii}\,$\lambda 1036$
line is less instructive in this context, partly because
of blending with \Lya\ forest lines.

%%%%%%%%%%%%%%%%%%
%%%  Figure 5  %%%
%%%%%%%%%%%%%%%%%%
\begin{figure*}
  \centering
  \includegraphics[angle=0,height=45.0mm]{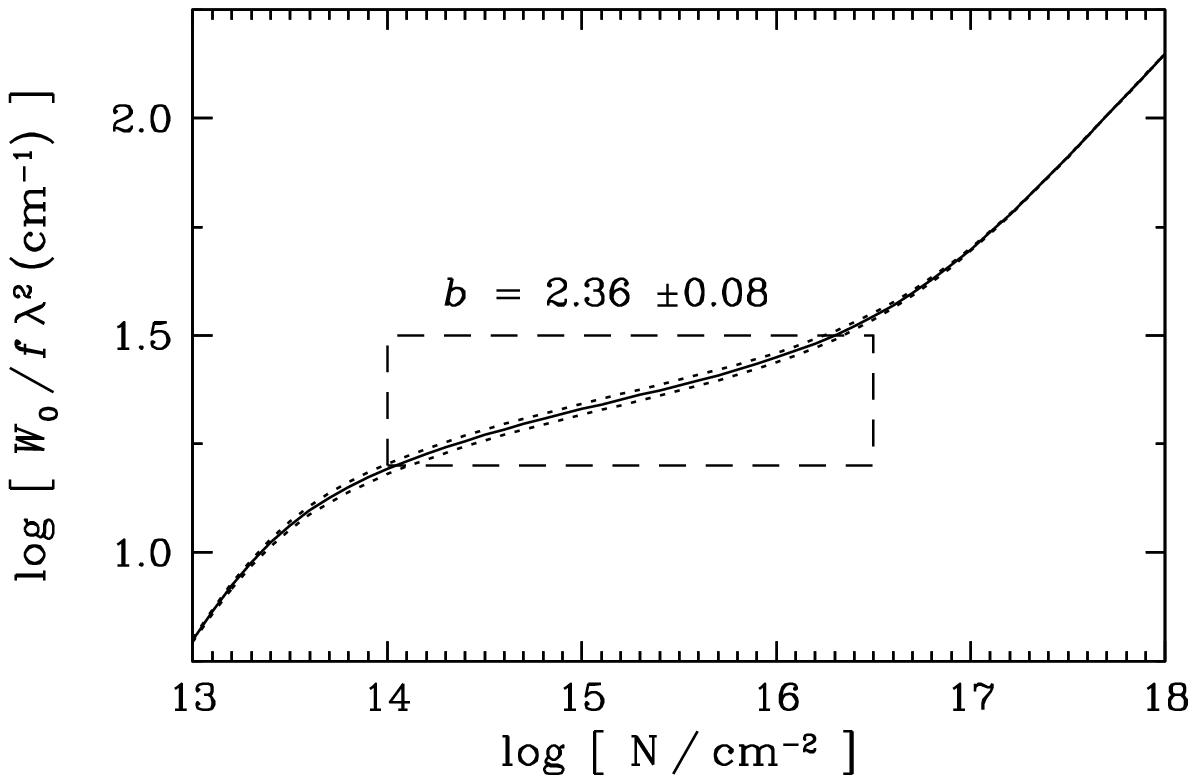}
  \includegraphics[angle=0,height=45.0mm]{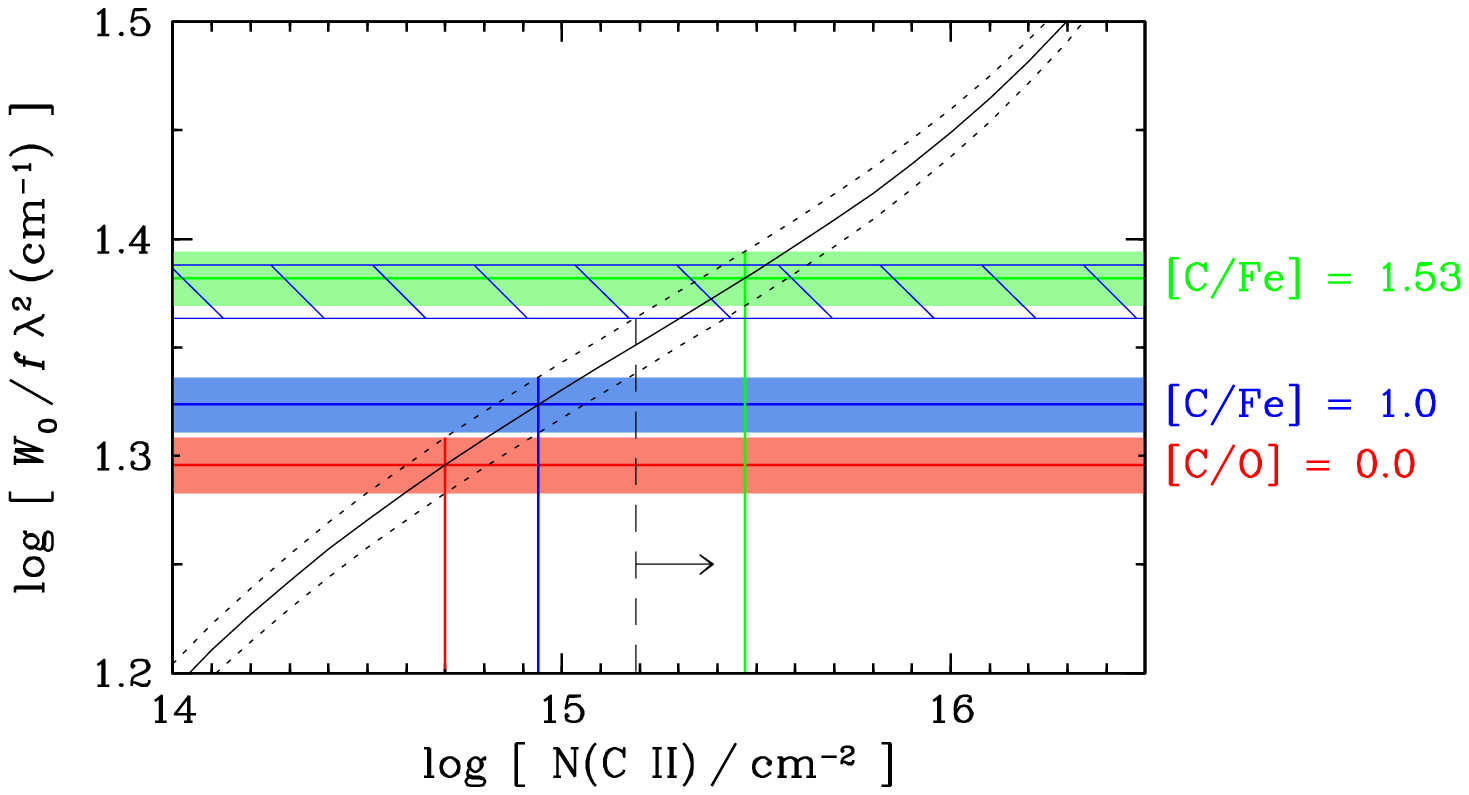}
  \caption{ 
\emph{Left panel}: Curve of growth for the $z_{\rm abs}=2.3400972$
DLA in line to J0035$-$0918, with Doppler parameter 
$b=2.36 \pm 0.08$\,km~s$^{-1}$. The dashed box shows the region
reproduced on an expanded scale in the right panel.
\emph{Right panel}: Equivalent width comparison.
The continuous black line is a portion of the curve of growth
for Doppler parameter $b=2.36$\,km~s$^{-1}$, with the short-dash
lines indicating the uncertainty $\delta b = \pm 0.08$\,km~s$^{-1}$.
The red, blue, and green horizontal bands show the 
corresponding ranges in equivalent
width $W_0$(C\,{\sc ii}$\lambda 1334$) 
appropriate to element ratios 
[C/O]\,=\,0.0, [C/Fe]\,=\,+1.0, and [C/Fe]\,=\,+1.53
respectively. Only the last of these 
(green band) matches the measured equivalent width
of C\,{\sc ii}$\lambda 1334$, indicated by the blue hatched 
horizontal band.
A strict lower limit [C/Fe]\,$\ge+1.25$ (vertical long-dash line
with right-pointing arrow) is obtained by considering
the lower bound on $W_0$(C\,{\sc ii}$\lambda 1334$)
and upper bound on $b$.
  }
  \label{fig:cog}
\end{figure*}

It may be surprising to some to see what subtle changes in
the line profiles result from changing $N$(C\,{\sc ii})
by a factor  of $\sim 6$. The reason is that the
two C\,{\sc ii} absorption lines are close to
saturation and, as they approach the flat part of the curve of growth,
their equivalent widths only
increase slowly with rising column density. 
In the present case, the curve of growth for the 
metal lines in the $z_{\rm abs} = 2.3400972$ DLA 
towards J0035$-$0918 (reproduced in Fig.~\ref{fig:cog})
is well-constrained by the relative 
strengths of six \OI\ and five \SiII\
transitions of widely differing $f$-values
(see Table~\ref{tab:lines}). 
Since the column densities of both \OI\ and \FeII\ are fixed
by optically thin transitions (i.e. independent of the
Doppler parameter), the ratios [C/O]
and [C/Fe] depend only on the \CII\ column density.
In the right panel of Fig.~\ref{fig:cog}, we compare
the measured equivalent width of C\,{\sc ii}\,$\lambda 1334$
(indicated by the blue hatched region)
with the values expected if 
[C/O]\,=\,0.0 (red), [C/Fe]\,=\,+1.0 (blue), and [C/Fe]\,=\,+1.53 (green). 
Only in this last case ([C/Fe]\,=\,+1.53) do we recover the measured
$W_0$(C\,{\sc ii}\,$\lambda 1334$), whereas the 
strength of the line is underpredicted in the other
two cases illustrated. 
The partially blended C\,{\sc ii}\,$\lambda 1036$ line is
not instructive in this equivalent width test which, by
its nature, is less sensitive than the pixel-by-pixel
line profile analysis illustrated in Fig.~\ref{fig:co_solar}.

\begin{figure}
  \centering
  \includegraphics[angle=0,width=72.5mm]{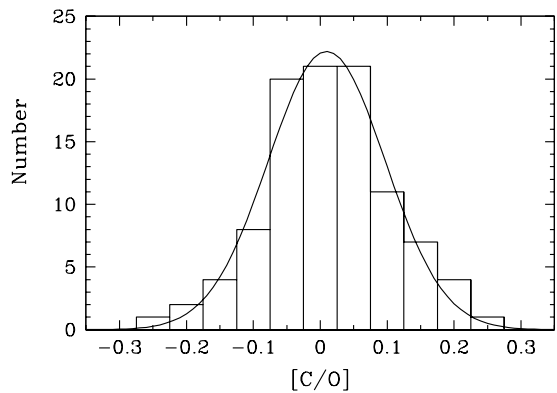}
  \caption{ 
Histogram of the values of [C/O] recovered from
100 Monte Carlo realizations with input [C/O]\,=\,0.0
in which the synthetic C\,\textsc{ii}\,$\lambda 1334$, $\lambda 1036$
line profiles were
perturbed with a random realization of the error spectrum.
The overlaid Gaussian fit to the results 
has a dispersion $\sigma = 0.087$, which implies 
that our measured value [C/O]\,$= +0.77$ 
would be a $\sim 9\sigma$ random fluctuation
if the [C/O] ratio were solar in the 
$z_{\rm} = 2.3400972$ DLA. 
  }
  \label{fig:mc_hist}
\end{figure}

Figures~\ref{fig:co_solar} and \ref{fig:cog} 
offer a clearer view of the carbon-enhanced nature
of this DLA, while at the same time highlighting 
the subtle differences in the absorption
line profiles when lines are on the flat part of the curve of growth.
Thus, it is reasonable to question whether the differences
between the line profiles computed with [C/O]\,=\,0.0
and the observed profiles, most evident for 
C\,{\sc ii}\,$\lambda 1334$ in the 
bottom right-hand set of panels in Fig.~\ref{fig:co_solar},
may be simply due to statistical fluctuations.
We tested this possibility with a Monte Carlo-type approach.
We synthesized 100 pairs of  \CII\,$\lambda 1036$, $\lambda$1334 line profiles, 
all corresponding to  [C/O]\,$ = 0.0$ and with the best fitting Doppler parameter 
($b = 2.36 \pm 0.08$). We perturbed these line profiles with Gaussian 
distributed errors from the observed $1 \sigma $ error spectrum, 
and then used \textsc{vpfit} to fit each of these 100 pairs of \CII\ lines 
(together 
with the observed line profiles of the other ions) 
and deduce 100 new values of $N$(\CII).
Using the simultaneously derived \OI\ column density 
(which is essentially unchanged since it is constrained by weaker transitions), 
we calculated the [C/O] ratio for all of our realizations to determine how often 
one would expect to observe [C/O] $\ge 0.77$ if the true [C/O] ratio were solar. 
The results of this Monte Carlo approach are shown in Fig.~\ref{fig:mc_hist}.
In none of the simulations was a value of [C/O] as high as $+0.77$
recovered. The distribution of recovered values is
symmetric about the input value [C/O]\,=\,0.0 with a dispersion
$\sigma = 0.087$. 
Thus, if the [C/O] ratio in the $z_{\rm} = 2.3400972$ DLA were
indeed solar (which in itself is higher or as high 
as observed in any other DLA up to now),
our data would constitute a  $\sim 9 \sigma$ deviation,
which is very unlikely indeed.

In conclusion, none of the tests we have performed can explain away
the C overabundance we have uncovered. 
The most straightforward interpretation of the data is that we have 
identified a damped \Lya\ system with a marked overabundance
of C, comparable to that so far found only in carbon-enhanced
metal-poor stars of the Milky Way.

%%%%%%%%%%%%%%%%%%%%%%%%%%%%%%%%%%%%%%%%%%%%%%%%%%%%%%%%%
\section{A Carbon-Enhanced metal-poor DLA}
\label{sec:discuss}
%%%%%%%%%%%%%%%%%%%%%%%%%%%%%%%%%%%%%%%%%%%%%%%%%%%%%%%%%

As already mentioned, this is the first example of a DLA meeting
the conventional definition of CEMP stars, [C/Fe]\,$> +1.0$.
While it is the only such case in the survey by \citet{Coo10b},
which includes a dozen DLAs with [Fe/H]\,$< -2$, it is nevertheless
very hard to draw conclusions regarding the frequency of 
CEMP DLAs. The reason stems from the difficulties
associated with measuring the \CII\ column density; 
the \CII\,$\lambda 1036$ and $\lambda 1334$ absorption lines are
almost always saturated in damped \Lya\ absorption systems, 
even at low metallicities.
CEMP DLAs may thus be easily overlooked.

%%%%%%%%%%%%%%%%%%
%%%  Figure 6  %%%
%%%%%%%%%%%%%%%%%%

The discovery of a gaseous, high redshift, counterpart
to CEMP stars of the Galactic halo is a breakthrough
with potentially very important consequences.
As mentioned in the Introduction, interpretations still
differ as to the origin of the carbon enhancement in very 
metal-poor stars, but the favoured explanation involves
mass transfer from an AGB companion, at least
for the members of the CEMP class which also exhibit
enhancements in $s$-process elements.
The situation is less clear-cut for the CEMP stars with no
such enhancement, the CEMP-no stars.
On the one hand, there may be
a continuous link between this class and 
the CEMP-s class \citep{Mas10}, so that
in some CEMP-no stars 
the carbon enhancement may also 
be due to mass transfer from an unseen companion.
On the other hand, 
the binany transfer scenario cannot readily
explain why the fraction of all
metal-poor stars that exhibit a carbon-enhancement
increases with decreasing metallicity \citep{BeeChr05}.
In fact, the only physical model so far put forward that
can explain this increased fraction \emph{requires}
a high carbon yield from metal-free stars to efficiently
cool low metallicity gas and drive the transition from
Pop~III to Pop~II star formation \citep{FreJohBro07}.

While CEMP-no stars in the Galactic halo 
may have more than one origin, 
the same ambiguities do not apply to the 
carbon enhancement in the DLA presented here, where
we have an entire cloud overabundant in carbon
(below we place limits to the total mass of carbon involved).
Rather, what we presumably see in the DLA is the initial chemical composition
of the gas from which CEMP stars would subsequently form.
In this case, the pattern of element abundances
revealed gives strong clues to the nature of the earlier generation
of stars (and supernovae) which seeded the DLA cloud with its metals.

We explore these ideas further by 
comparing the abundance pattern of the DLA
with those of halo stars of similar metallicity and with 
calculations of element yields from metal-free and metal-poor
stars. In this endeavour,
we are limited by the small number
of elements whose abundances we have been able to measure, 
but this is an inevitable consequence of the low metallicity
of this DLA. While it would be instructive to know the
abundances of other Fe-peak elements, such as Cr and Zn for
example, the absorption lines of these species would require
S/N\,$ \simgt 500$ for  $5 \sigma$ detections!\footnote{Assuming
[Cr/Fe] = [Zn/Fe] = 0.0\,.}
Realistically, only Mg and S are within reach of dedicated
future observations, and there is certainly no
hope of measuring directly the
abundances of neutron capture elements in the DLA.

\subsection{Comparison with Stellar Abundances}

In order to compare the relative abundances of the elements
measured in the DLA with the corresponding values in 
Galactic halo stars,
we median-combined the abundances of all CEMP stars 
with  Fe/H within a factor of two of the
DLA ($-3.34\le{\rm [Fe/H]}\le-2.74$) 
from the recent compilation of \citet{Fre10}.
We then calculated the dispersion in each abundance using the
Interactive Data Language routine
\textsc{robust\_sigma}\footnote{Available from http://idlastro.gsfc.nasa.gov/homepage.html}
which determines the median absolute deviation
(unaffected by outliers) of a set of measurements,
and then appropriately weights the data to provide a
robust estimate of the sample dispersion (\citealt{HoaMosTuk83}, cf. \citealt{Coo10a}).

%%%%%%%%%%%%%%%%%%
%%%  Figure 7  %%%
%%%%%%%%%%%%%%%%%%
\begin{figure}
  \centering
  \includegraphics[angle=0,width=80mm]{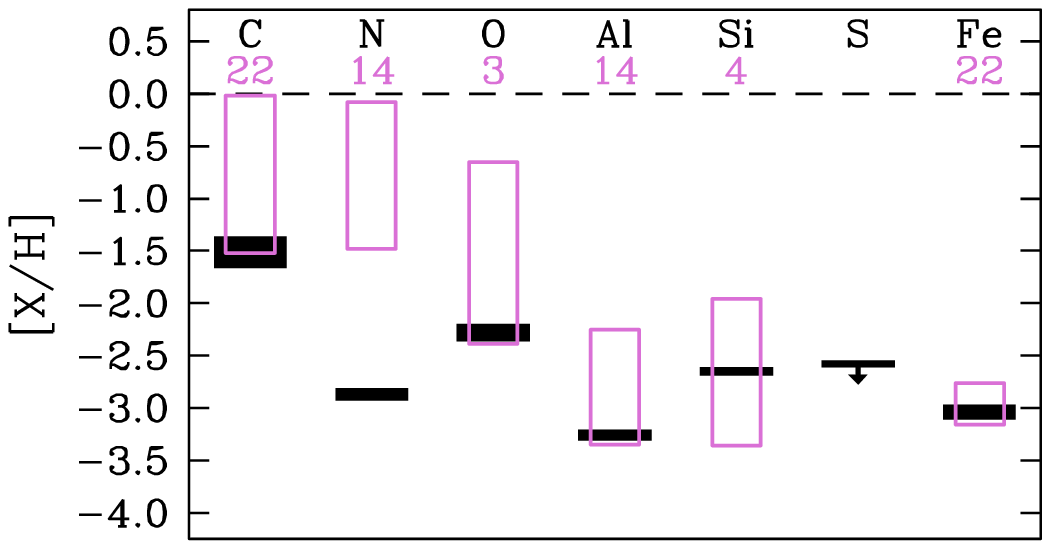}
  \includegraphics[angle=0,width=80mm]{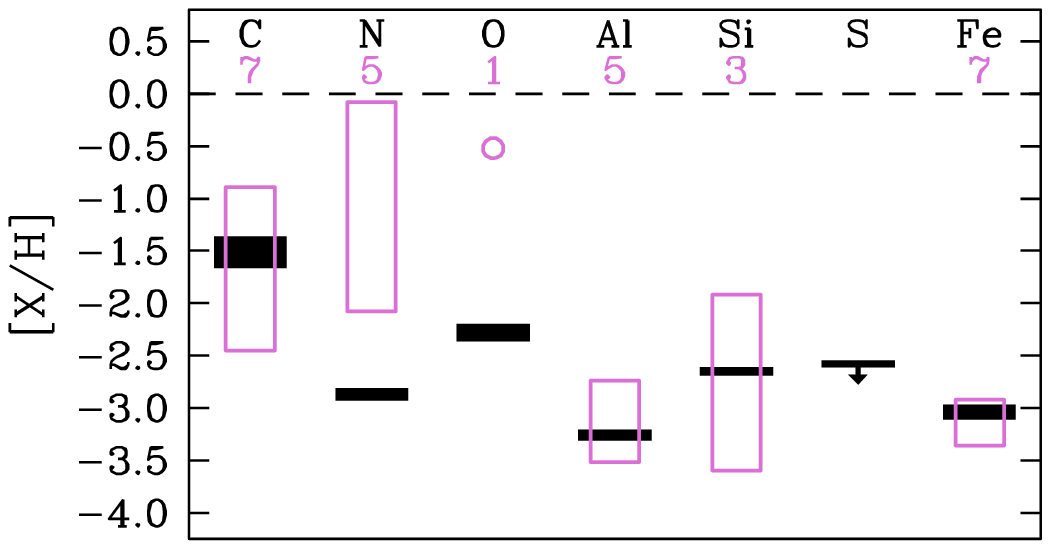}
  \includegraphics[angle=0,width=80mm]{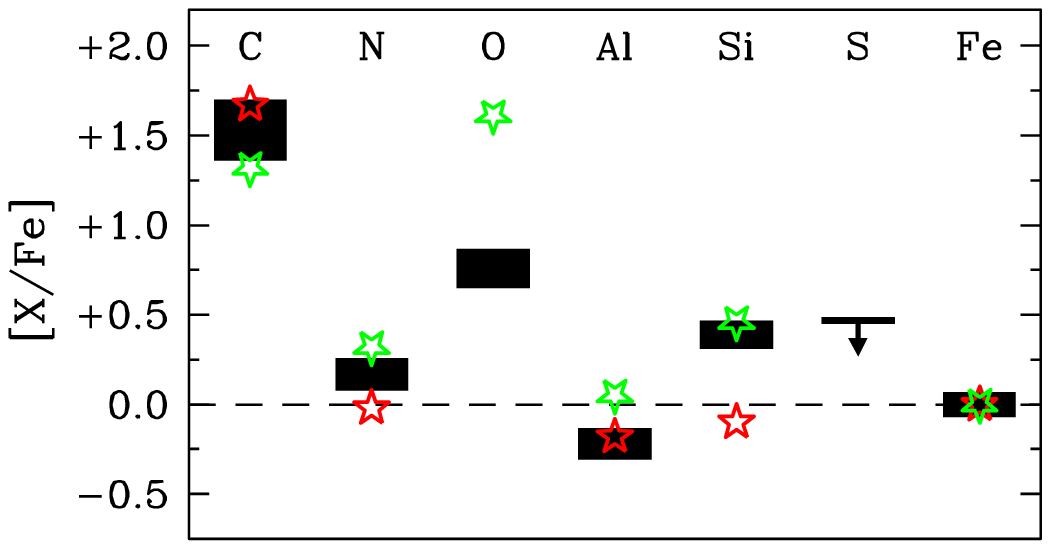}
  \caption{
  Comparison of element abundances in the $z_{\rm abs}=2.3400972$ DLA
  (filled black boxes) 
  and in Galactic halo stars with an Fe abundance within a factor
  of two of the DLA (open magenta boxes).
  The numbers below the element labels indicate the number 
  of stars that contributed to the 
  determination of the `typical' stellar abundances,
  and the heights of the magenta boxes reflect the dispersion of each set
  of measurements.
  \textit{Top Panel:} comparison with all CEMP stars
  that have $-3.34\le{\rm [Fe/H]}\le-2.74$.
  \textit{Middle Panel:} comparison with CEMP-no stars
  that have $-3.34\le{\rm [Fe/H]}\le-2.74$.
  For this case, the oxygen abundance of a 
  single CEMP-no star is shown by the open circle.
  \textit{Lower Panel:} comparing the DLA abundance pattern with
  the stellar abundance patterns of
  HE\,0143$-$0441 (red symbols; a CEMP-s star with 
  [Fe/H]\,$ = -2.21$, [Ba/Fe]\,$ = +0.62$ from \citealt{Coh04}) and
  BD$+44^{\circ}493$ (green symbols; 
  a CEMP-no star with [Fe/H]\,$=-3.73$, [Ba/Fe]\,$=-0.55$
  from \citealt{Ito09}).
  Note that in this last panel, we have plotted [X/Fe] as opposed to [X/H].
  In all panels the dashed line represents the solar abundance.
 }
  \label{fig:abund_plot_mpstars}
\end{figure}

Figure~\ref{fig:abund_plot_mpstars} illustrates three comparisons.
In the top panel, the DLA abundance pattern (solid black boxes)
is shown together with that of the
`average' population of all CEMP stars (open magenta boxes),
while in the middle panel the comparison is restricted to CEMP-no stars.
Clearly CEMP stars, of both flavours, are a heterogeneous population,
exhibiting wide ranges in the ratios of the elements considered here.
On the basis of the limited evidence available, it appears that 
the C/Fe ratio in our DLA is a better match to the values seen
in the subset of CEMP-no stars than to the more extreme values 
encountered in some CEMP stars.
Other elements ratios are less straightforward to interpret.

Nitrogen, for example, is evidently less abundant in 
the DLA compared to CEMP stars
by $\sim 2$ orders of magnitude!
However, it is difficult to draw firm conclusions
from this discrepancy, because the
abundance of N is  
notoriously difficult to measure in metal-poor stars
where it relies on the analysis of 
molecular bands of NH or CN,
each with its own disadvantages (see e.g. \citealt{Asp05}).
The largest uncertainty arises from the assumed 1D (rather than 3D) 
model atmospheres; correcting for 3D effects can change
then deduced values of N/H by $\sim -0.5$ dex
(Note that similar corrections also apply to C/H
when derived from molecular bands of CH or C$_{2}$; \citealt{ColAspTra07}).
Furthermore, the 
photospheric  N/H we measure today is
unlikely to reflect the 
N abundance of the gas cloud that
gave birth to the star; during the star's
evolution it will become self-polluted by its
own nucleosynthesis (e.g. through CN cycling
or rotational mixing), which is likely emphasised by
the high seed C abundance. Thus, a comparison between
stellar and DLA nitrogen abundances is not very instructive.

Comparing the oxygen abundances in CEMP stars and
the DLA is similarly problematic. The statistics
are very poor  (see Fig.~\ref{fig:abund_plot_mpstars})
and it is well known that different spectral features
used to deduce O/H in the metal-poor regime give
discordant answers \citep{Gar06}.
Significant departures from local thermodynamic equilibrium (LTE)
become important when measuring the O abundance
from the infrared triplet at $777$ nm in the metal-poor regime 
([Fe/H]\,$<-2.5$), 
leading to non-LTE corrections of the order of
$-0.5$ to $-1.0$ dex \citep{Fab09b}.
Similarly, the oxygen abundance determined from the
UV OH lines is subject to large 3D corrections of up to $-0.9$ dex
when [Fe/H]$\,\le-3.0$ \citep{Asp05}.
Clearly, it will be important to reexamine this issue once
samples of CEMP stars have grown to include
more cases where [O/Fe] has been measured with confidence.

Finally, Fig.~\ref{fig:abund_plot_mpstars} 
demonstrates that there is a reasonably good match between 
CEMP stars and the DLA in the elements Al and Si, 
particularly for the CEMP-no stars.
In this context, we point out that significant 
(positive) non-LTE corrections to Al/H may 
apply for stars in this metallicity regime 
(see e.g. \citealt{Asp05}). To address this 
issue, we first identified those CEMP stars
in the \citet{Fre10} compilation whose Al abundances
were derived assuming LTE.
We then applied non-LTE corrections to
Al based on the \citet{And08} estimates,
which amounted to a typical correction of the order of $+0.6$.
The Al abundances for the `typical' CEMP and 
CEMP-no star presented
in Fig.~\ref{fig:abund_plot_mpstars} should thus correspond 
to the (approximate) non-LTE values.

Finally, in the lower panel of Fig.~\ref{fig:abund_plot_mpstars}
we compare the element ratios (relative to Fe) in the DLA with the 
values measured for two CEMP stars selected from the compilation
by  \citet{Fre10} because they most closely match the relative abundances
in the DLA. It is certainly plausible that stars forming out of this
DLA gas would share many chemical similarities -- at least for
the elements considered here -- with these two stars of the
Milky Way Galactic halo.

\subsection{Comparison with Model Yields for Metal-free Stars}

With a metallicity [Fe/H] of only 1/1000 solar, it is conceivable
that the chemical composition of the DLA gas reflects the element
yields of only a few prior generations of stars. It is thus of interest
to compare the abundance pattern in Fig.~\ref{fig:abund_plot} with
calculations of nucleosynthesis by metal-free, or low metallicity, stars.
Calculations of yields from metal-free stars (commonly referred to as 
Pop\,III stars), aimed in particular at interpreting the element ratios
seen in very metal-poor stars, have focused on models which include 
`mixing and  fallback' \citep[e.g.][]{UmeNom02,UmeNom03,HegWoo08}.
These scenarios involve core-collapse supernovae where the elements
synthesized in the inner regions of the star are mixed by some
process (several possibilities have been put forward); a fraction
of the mixed material subsequently falls back onto the central 
remnant while the rest is ejected into interstellar space.

Such models are illustrative, rather than predictive,
since the parameters describing the boundaries of the mixing region
and the fraction of the mixed material which is ejected cannot be
derived from first principles, and are instead parameterised and suitably adjusted to
fit the observed stellar abundances (e.g. \citealt{TomUmeNom07}).
A string of recent models that employ a more physically motivated
prescription of fallback \citep{HegWoo08},
have also investigated the effects of mixing
due to the Rayleigh-Taylor instability \citep{JogWooHeg09} together 
with rotationally induced mixing in 
2D \citep{Jog10a} and 3D \citep{Jog10b}.
By and large, a general feature of these models is that an
abundance pattern similar to that found here in the DLA,
including the carbon enhancement and the marked odd-even effect
(see Fig.~\ref{fig:abund_plot}),
can be reproduced with low energy explosions, 
such as those giving rise to the faint SN branch
(see Fig.~1 of \citealt{TomUmeNom07}, and Fig.~11 of \citealt{HegWoo08}),
and a moderate degree of mixing and fallback.
For example, in the mixing-fallback
model considered by \citet{Tom10},
enrichment with the elements 
ejected by a single $25\,M_\odot$ 
star of zero metallicity, exploding 
as a core-collapse supernova with an 
energy $E_{\rm SN} \sim 10^{51}$\,ergs,
provides a remarkably good fit to the relative element
abundances measured in the DLA (Kobayashi et al. in prep).

It remains to be established whether a zero
initial metallicity (i.e. Pop~III)
is actually required to match
the observations, or whether a similar abundance
pattern could also be reproduced by models with low, 
but non-zero, initial metallicity.
An alternative to the mixing and fallback model
has been proposed by the Geneva group
whose work places more emphasis on the effects 
of rotation to provide the mixing and the trigger
for mass loss in very metal-poor stars
\citep[e.g.][]{MeyEksMae06,Hir07,Mey10}.
Whilst these models are physically well-motivated, the
rather high nitrogen yield predicted by some models contrasts
with the relatively low [N/C] and [N/O] observed in this DLA.
This may in turn place interesting contraints on the rotation
velocities of the metal-free stars that may have 
seeded the DLA with its metals.
Only recently has rotation been included in zero-metallicity
progenitors under the mixing-fallback scenario \citep{Jog10a}.
% For the different rotation rates considered by these authors,
% the final \emph{structure} of the star is relatively unchanged; rotational
% mixing seems to be less important. Introducing even a modest rotation rate,
% however, changes the zero-metallicity progenitors from a compact blue star 
% to a red giant, which in turn indirectly affects the final yield. 
% The reverse 
% shock during the explosion takes longer to propagate through the now larger star,
% meaning that the pressure gradient is reversed for a longer time,
% which causes the stellar layers to undergo \emph{more} mixing via
% the Rayleigh-Taylor instability. 
It should be possible to assess better
the effects of rotation on the yield from zero-metallicity progenitors
as these stellar models improve and as the samples of CEMP DLAs grow.

\subsection{How Many Supernovae?}

It is truly intriguing that an \emph{entire} cloud of neutral gas is
so highly enhanced in carbon and shows such a clear-cut 
odd-even effect, if both are indeed
characteristic of Pop~III supernova yields.
In this context, it is of interest to estimate the 
mass of carbon involved. 
With the assumption of spherical symmetry, 
the mass of singly ionized carbon in the DLA is given by:
\begin{eqnarray}
M({\rm C\,\textsc{ii}}) \!\!&=&\!\! 12 \, M({\rm H\,\textsc{i}}) \, \frac{N({\rm C\,\textsc{ii}})}{N({\rm H\,{\textsc{i}}})} \\
M({\rm C\,\textsc{ii}}) \!\!&=&\!\! 2 \, \pi \, m_{\rm H} \frac{N({\rm H\,\textsc{i}})^{2}N({\rm C\,\textsc{ii}})}{n({\rm H})^{2}} \\
M({\rm C\,\textsc{ii}}) \!\!&\leq&\!\! 200 \, \bigg( \frac{n({\rm H})}{0.1\,{\rm cm}^{-3}} \bigg)^{-2} \, M_{\odot}
\end{eqnarray}
where $m_{\rm H}$ is the mass of a hydrogen atom and
$n$(H) is the volume density (cm$^{-3}$).
The corresponding mass of neutral gas in the DLA is:
\begin{eqnarray}
M_{\rm DLA} \!\!&=&\!\! 1.3 \, m_{\rm H} \, \frac{4\pi}{3} \frac{N({\rm H\,\textsc{i}})^{3}}{8\,n({\rm H})^{2}} \\
M_{\rm DLA} \!\!&\leq&\!\! 2.5 \times 10^{6} \, \bigg( \frac{n({\rm H})}{0.1\,{\rm cm}^{-3}} \bigg)^{-2} \, M_{\odot} 
\end{eqnarray}

Since C is mostly singly-ionized in DLAs, eq.~(4) implies an upper limit
of $200\,M_\odot$ to the total mass of $^{12}$C in the DLA. 
However, the inverse square
dependence of $M$(C\,\textsc{ii}) on the gas density 
allows substantially lower values of $M$(C\,\textsc{ii}). 
In Section~\ref{sec:ion_corr} it was shown that our lower limit
$n{\rm (H)} \geq 0.1$\,cm$^{-3}$ was based on the non-detection
of N\,\textsc {ii} absorption, and that higher densities 
are likely given the very low velocity dispersion of the gas.
If, for example, $n{\rm (H)} \simeq 1$\,cm$^{-3}$, which would
imply a linear extent of the DLA of 100\,pc along the line
of sight, then the implied mass of $^{12}$C would be reduced to only 
$2\,M_\odot$. For comparison, the total mass of $^{12}$C ejected by the
single $25\,M_\odot$ Pop~III star in the model by \citet{Tom10}
is $\sim 0.2 M_\odot$. The models by \citet{HegWoo08}
anticipate a similar ejected mass of $^{12}$C ($\sim0.3\,M_{\odot}$)
for a comparable explosion energy and progenitor mass. 
Thus, we may indeed be seeing the elements synthesized
by only a few supernovae in the chemical enrichment of the
DLA considered here.

In conclusion, we speculate that the $z_{\rm abs}=2.3400972$ DLA
in front of the QSO J0035$-$0918 may well be the much sought
`missing link'
between the first, zero-metallicity, stars 
and the most metal-poor stars in the halo of our Galaxy. 
Its low metallicity of 1/1000 solar in Fe, coupled with 
an overabundance of C and a marked odd-even effect
in the relative abundances of the few elements that could
be measured, are all consistent with the yields produced 
by models of Pop~III stars which explode as core-collapse
supernovae of relatively low energy. The mass of newly synthesized
elements may be that produced by only a few such supernovae,
depending on the unknown volume density of the gas in the DLA.
In this scenario, the gas we see as a damped \Lya\ system at
high redshift may be the material from which a subsequent
generation of stars formed, with a chemical composition similar to that
seen in CEMP-no stars of the Galactic halo.

\section{Summary and Conclusions}
\label{sec:conc}

In the course of our program to study the most metal-poor DLAs and,  
in particular, to measure the abundances of the CNO group of elements, 
we have uncovered a DLA, at $z_{\rm} = 2.3400972$
towards the $z_{\rm em} = 2.42$ SDSS QSO J0035$-$0918, 
whose chemical composition
is consistent with that produced from exploding Population\,III stars.
From the analysis of medium and high resolution spectra of this
DLA we draw the following conclusions.

\smallskip

\noindent ~(i) The metal absorption lines associated
with the DLA are formed in quiescent gas, consisting
of a single absorption component with a small
Doppler parameter $b = 2.4 \,{\rm km\,\,s}^{-1}$.

\noindent ~(ii) The metallicity of the DLA, as measured from Fe,
is very low: [Fe/H]\,$= -3.04 \pm 0.17$, or $\sim 1/1000$ solar.

\noindent ~(iii) The DLA exhibits a strong enhancement of the 
abundance of carbon relative to all other
metals covered by our data: N, O, Al, Si, S and Fe. 
We measure [C/Fe]\,$= +1.53$, a factor of $\sim 20$
% (or $\sim1.3\,\,{\rm dex}$) 
greater than observed in any other DLA up to now.
Adopting the defining criterion for carbon-enhanced metal-poor stars in 
the Galactic halo, [C/Fe] $> +1.0$,  this is
the first example of an analogous carbon-enhanced DLA.
We also deduce [C/O]\,$ = +0.77$, whereas in all other
DLAs with [Fe/H]\,$<  -2$ studied up to now [C/O]\,$ \simlt 0.0$.

\noindent ~(iv) The DLA also exhibits a clear odd-even effect,
which implies a low neutron excess, and hence presumably
low abundances of neutron-capture elements.
When its chemical composition is compared with that
of Galactic carbon-enhanced metal-poor stars which do not
exhibit an excess of neutron-capture elements, a good match is
found for some element ratios (C/Fe, Al/Fe and Si/Fe).
N and O are significantly less abundant, compared to Fe, 
in the DLA than in most CEMP stars, but it is difficult to draw
definite conclusions for these two elements which are
notoriously difficult to measure in very metal-poor stars.

\noindent ~(v) The abundance pattern we observe for this DLA is
consistent with enrichment from a population 
of $\sim 20$--$50\,{\rm M}_{\odot}$ metal-free,
or extremely metal-poor, stars that ended
their lives as core-collapse supernovae with modest explosion energies.
We estimate 
the total mass of $^{12}$C within the DLA
to be $\leq 200\,\,M_{\odot}$. This upper limit 
could be constrained further with higher signal-to-noise ratio spectra 
that would permit a measure of the column density of \NII\ and other
successive ion stages whose ratios are density dependent.
The steep dependence of $M$($^{12}$C) on the gas density
allows the possibility that we may be seeing the chemical enrichment
produced by only a few prior supernovae.

\noindent ~(vi) We speculate that the gas in this DLA may 
be the `missing link' between
the yields of Population~III stars, and their later incorporation in the
CEMP-no class of carbon-enhanced metal-poor stars. 
We note, however, that the carbon-enhancement in 
some CEMP-no stars could also be
produced by other means.
Long term radial velocity monitoring of CEMP-no stars 
will confirm or deny
their association with a now extinct companion.

The results presented here emphasise the importance of 
further observations of DLAs with [Fe/H]\,$\leq  -3$ 
to complement the work being carried out on the most
metal-poor stars. Such DLAs may be the 
most suitable environments for measuring the 
true yields from zero- or low-metallicity stars,
free from the complications of stellar abundance
measurements and the possibility that these stars
are polluted by an
unseen binary companion,
or self-polluted by their own nucleosynthesis.
The extremely metal-poor regime for DLAs is yet to
be thoroughly
explored with high-resolution and high signal-to-noise 
spectra---who knows what may lurk there!

\section*{Acknowledgements}
We are grateful to the staff at the Keck and Magellan telescopes
for their assistance with the observations,
and to Tom Barlow, George Becker, Bob Carswell, Gary Ferland,
and Michael Murphy for
providing much of the software used in the reduction
and analysis of the data.
We wish to thank Poul Nissen and Chris Tout
for useful discussions regarding the stellar abundances.
Chiaki Kobayashi and Nozomu Tominaga
kindly helped with the interpretation
of the results.
A prompt and comprehensive referee report significantly
improved the paper.
We thank the Hawaiian
people for the opportunity to observe from Mauna Kea;
without their hospitality, this work would not have been possible.
RC is jointly funded by the Cambridge Overseas 
Trust and the Cambridge Commonwealth/Australia Trust 
with an Allen Cambridge Australia Trust Scholarship.
CCS's research is partly supported by grants
AST-0606912 and AST-0908805 from the US National Science Foundation.

%\bsp

\label{lastpage}

\end{document}